\documentclass[aps,amsmath,amssymb,prl,twocolumn,showpacs,superscriptaddress]{revtex4}

\usepackage{amsfonts}
\usepackage{graphicx}
\usepackage{color}

\begin{document}

\title{Current Correlations from a Mesoscopic Anyon Collider}

\author{Bernd Rosenow}
\affiliation{Institut f\"ur Theoretische Physik, Universit\"at Leipzig, D-04009 Leipzig, Germany}
\affiliation{Physics Department, Harvard University, Cambridge, Massachusetts 02138, USA}
\author{Ivan P. Levkivskyi}
\affiliation{Theoretische Physik, ETH Zurich, CH-8093 Zurich, Switzerland}
\affiliation{Physics Department, Harvard University, Cambridge, Massachusetts 02138, USA}
\author{Bertrand I. Halperin}
\affiliation{Physics Department, Harvard University, Cambridge, Massachusetts 02138, USA}

\date{\today}

\begin{abstract}

Fermions and bosons are fundamental realizations of exchange statistics, which governs the probability for two particles being close to each other spatially. Anyons in the fractional quantum Hall effect are an example for exchange statistics intermediate between bosons and fermions. We analyze a mesoscopic setup in which two dilute beams of anyons collide with each other, and relate the correlations of current fluctuations to the probability of particles excluding each other spatially. While current correlations for fermions vanish, 
negative correlations for anyons are a clear signature of a reduced spatial exclusion as compared to fermions.

\end{abstract}


\maketitle

One of the important differences between fermions and bosons is the difference in the probability of two identical particles being close to each other, which is associated with their exchange statistics. These differences are manifest in various ways, including Hanbury Brown-Twiss interference experiments \cite{HBT54,HBT56}. In the integer quantum Hall (IQH) regime, such two-particle interferometers have been experimentally realized 
\cite{Neder+07}.  It is tempting to see whether any of these distinctions can be carried over to particles with fractional statistics such as appear in the fractional quantum Hall (FQH) effect  \cite{LeMy77,Laughlin83,Halperin84,ArScWi84}.

\begin{figure}[t]
\includegraphics[width=0.9\linewidth]{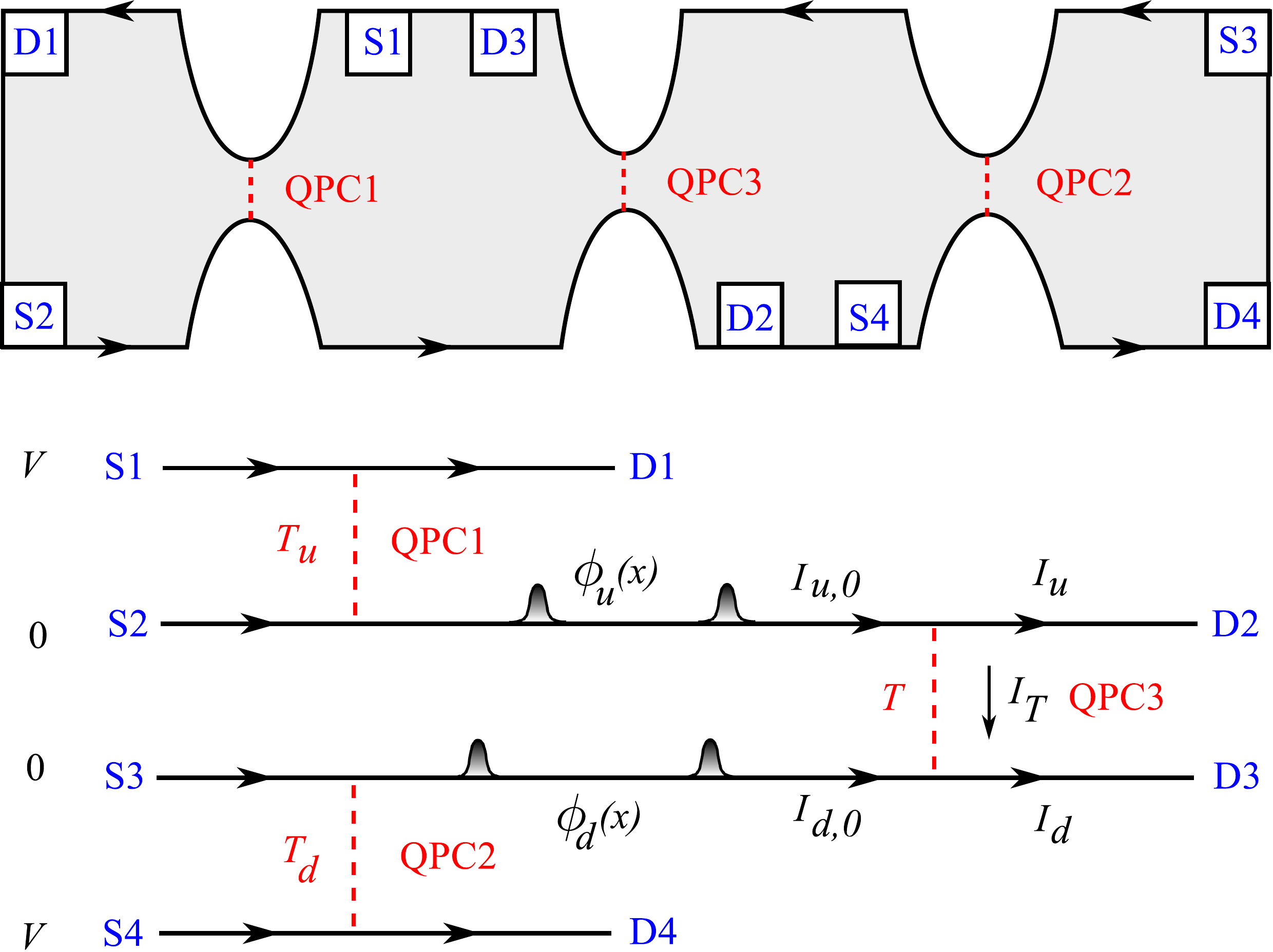}
\caption{Sketch of the anyon collider, with Hall bar geometry in the upper panel  and idealized geometry in the lower panel. Dilute quasi-particle beams are generate at QPC1 and QPC2, propagate along upper (u) and lower (d) edge, and collide at QPC3.  Sources S1 and S4 are at voltage $V$, sources S2 and S3 are grounded.  }
\label{fig:setup}
\end{figure}

A possible realization of an anyonic two-particle interferometer was suggested by Campagnano 
et al.~\cite{Campagnano+12}. There, it was found that correlations  exhibit partial bunching similar to bosons, but 
there also exist qualitative differences between the anyonic signal and the corresponding bosonic or fermionic signals. Here,  we are proposing and analyzing a somewhat simpler experiment which does not depend on interference loops or phase coherence. We consider specifically the anyons which occur at the clean chiral edge of a quantized Hall state \cite{Halperin82,Wen90}. We discuss explicitly only the single edge mode case but discuss the generalization to multi-mode cases qualitatively.
In the  setup Fig.~\ref{fig:setup},  two dilute beams of anyons are produced on the upper and lower edge, which are
eventually connected by a quantum point contact (QPC). We are considering current correlations on the two edges downstream from the QPC, and study cross-correlations at low frequencies, low temperatures and low voltages, such that the details of interactions at short distances are not important for the final results. 

%
%*********************************  setup   **************
%***************************************************************
%
 For non-interacting fermions as occurring on the edge of the IQH effect \cite{Halperin82,Landauer70,Buttiker88}, we find that cross-correlations are absent at zero net bias between the two edges, whereas for anyons we find finite cross-correlations. One may attempt to characterize these results as a measure of exclusion statistics. The results for the cross-correlations are related to the power laws governing the long time decay of the correlation functions on the edge, which 
due to conformal invariance are related to the spactial dependence which reflects the braiding statistics of anyons in the bulk. 

\begin{figure}[t]
\includegraphics[width=\linewidth]{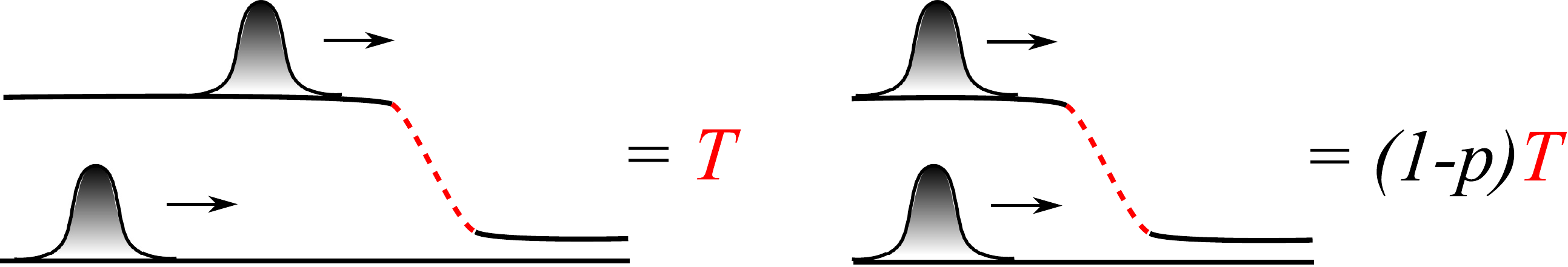}
\caption{Definition of the exclusion probaility $p$. Two particles arriving on opposite edges have a reduced probability $(1-p)T$ to continue together on the same edge 
after tunneling with a small single particle probability $T$ at a QPC. }
\label{fig:exclusion}
\end{figure}

In order to get some intuition for the results of our quantum mechanical calculation, we compare with  the results of a classical 
lattice model with a two-particle exclusion probability $p$. For fermions, $p=1$ indicates that two particles can never occupy the same spatial position, 
whereas $p=0$ would describes  the absence of spatial correlations.  The
probability that two anyons incident on different edges continue to propagate on the same edge after the collision is suppressed by a factor  $1-p$ if they occupy the same position in space, see Fig.~\ref{fig:exclusion}.  Although there are similarities on a 
qualitative level between the quantum mechanical and the classical model, we find that the quantum mechanical result has a level of universality not present in the classical calcuation, such that a quantitative comparison is difficult. 

We consider a mesoscopic collider for anyons, in which diluted beams  are created with the help of quantum point contacts QPC1 and QPC2 with small tunneling probabilities $T_u$ and $T_d$, see Fig.~\ref{fig:setup}. At a third QPC3 with tunneling probability $T$, the  beams collide with each other. Statistical fluctuations  are reflected in the cross-correlation  
$\langle \delta I_u \delta I_d\rangle_{\omega=0} $ at zero frequency between current fluctuations 
after the collision. Indeed, for the case of two incoming beams with  equal magnitude of curent,  we find $\langle \delta I_u \delta I_d\rangle_{\omega=0} = 0 $ for the case of fermions, while for a lattice model of non-interacting particles with a general exclusion 
probability $p$, we find $\langle \delta I_u \delta I_d\rangle_{\omega=0} \propto -(1-p) $, see \cite{suppmat}, indicating that current cross-correlations contain 
information about the exclusion proabability.

A fully quantum mechanical description 
needs to take into account that an anyon which tunnels into a fractional quantum Hall edge is dressed by charge density fluctuations, and that the 
anyonic correlation function decays like a power law with time and distance. Due to the slow decay of the correlation function, 
an approach which is perturbative in the weak tunneling probability of QPC1 and QPC2  fails \cite{KaFi03}, 
and  the non-pertubative method of non-equilibrium bosonization \cite{LeSu09,GuGeMi}   as adapted to fractional qps \cite{Levkivskyi14}   is needed.   For the case of  quantum Hall states from the Laughlin series with filling fraction $1/m$ with odd $m$, 
we find $\langle \delta I_u \delta I_d\rangle_{\omega=0} 
\propto -2/(m-2)$. This result is clear evidence for a reduced exclusion probability $p <1$ as compared to fermions.
For $m=3$, a comparison between the result for anyons and that for the lattice model suggests $p < 0$,  and hence evidence for bunching of particles. Interestingly, for larger $m$ cross-correlations are reduced, despite the fact that the statistical angle $\theta = \pi/m$ 
approaches the bosonic value $\theta = 0$, which would naively suggest an increase of bunching.  We demonstrate that the reduction of cross-correlations for increasing $m$ is due to the fact that the anyonic non-equilibrium state increasingly resembles  a thermal state with vanishing  current cross-correlations. 

%
%*********************************  full noise minussign   **************
%***************************************************************
%

{\em Free fermions:} the current cross-correlations can be computed by using a scattering description of  the QPC3 connecting upper and lower wire. We use an S-matrix $S = \exp(i \gamma \sigma_y)$ to connect the outgoing fermions $\Psi_\alpha$ with the incoming 
ones $\Psi^{(0)}_\alpha$, with $\sigma_y$ denoting the corresponding Pauli matrix, and $T = \sin^2 \gamma$ denoting the probability for tunneling across the QPC . Here, $\alpha = u, d$ parametrizes upper and lower edge, respectively. Using the fact that for chiral fermions the current is proportional to the density $\Psi^\dagger_\alpha \Psi_\alpha$, the current cross-correlations of the out-going fermions can be related to the distribution functions $f_\alpha(\epsilon)$ of the incoming ones via \cite{BlBu00}
%
%**********************  free fermion cross-correlation ***********
\begin{eqnarray}
\langle \delta I_u \delta I_d\rangle_{\omega=0} \! & = & \! T(1\! - \!T)  {e^2 \over h}\!  \int \!  \! d\epsilon \left\{ f_u(\epsilon)[1\! -\! f_u(\epsilon)] 
\! 
\right. \label{fermion_correlation.eq} \\
&  & \hspace*{-2cm} \left. +\! f_d(\epsilon)[1\! -\! f_d(\epsilon)] - 
f_u(\epsilon)[1-f_d(\epsilon)] - f_d(\epsilon)[1-f_u(\epsilon)]\right\}  . \nonumber
\end{eqnarray}
%****************************************************
% 
If both $f_u$ and $f_d$ are Fermi distributions at the same temperature and chemical potential, cross-correlations clearly vanish.  
Applying a  non-equilibrium current bias to the incoming fermions via  QPC1 and QPC2, which are connected to source contacts at a bias voltage $V$, the incoming fermions have double step distributions 
$f_\alpha(\epsilon) = \theta(-\epsilon) + T_\alpha \, \theta(\epsilon) \theta(V - \epsilon)$. 
Then, $ \langle \delta I_u \delta I_d\rangle_{\omega=0} = - T (1-T)V(e^2/h)(T_u - T_d)^2$, and we see that for equal bias currents with $T_u =  T_d$ the cross-correlations vanish. The absence of cross-correlations in the limit of zero effective bias is due to a cancellation between the first two terms in Eq.~(\ref{fermion_correlation.eq}), which describe the effect of partially transmitted fluctuations in the incoming currents, and the last two terms, which describe noise generated at the QPC.

{\em Classical lattice model:} in order to link the absence of cross-correlations in the case of zero net bias, $T_u=T_d$, to quantum statistical properties of fermions, 
we now analyze a lattice model, where both upper and lower edge are described by a one-dimensional chain. Initially, each lattice site is occupied with  probabilityies 
$T_{u,d} \ll1$. In each time step, particles move one site forward. When a  particle moving along the upper edge arrives at the lattice site representing QPC3, and if no particle arrives at the same time on the lower edge, 
the particle tunnels from the upper to the lower edge or vice versa with probability $T$. However, when particles arrive at QPC3 simultaneously on both the upper and the lower edge, the probability for tunneling is reduced to $(1-p) T$ in an analysis to leading order in $T$. Thus, the probability for both particles continuing on the same edge is $(1-p) 2 T$, and the probability for continuing on opposite edges is $1 - (1-p)2 T$. We identify the parameter $p$ as the  exclusion probability introduced in the introduction: fermions are sure to exclude each other corresponding to $p=1$, and for  general $p$,  we find the result \cite{suppmat}
%
%*********************   classical model cross-correlations  *****************
\begin{equation}
\langle \delta I_u \delta I_d\rangle_{\omega=0} \ = \ - T V{e^3 \over h}(T_u - T_d)^2 - 2 T V{e^3 \over h}(1-p)T_u T_d \ .
\label{classical_correlation.eq}
\end{equation}
%*****************************************************************
%
For $p=1$, we exactly reproduce the fermionic formula derived above at small $T$. For general $p$, 
and for vanishing net bias with $T_u = T_d$, cross-correlations are proportional to $1-p$, i.e.~to the deviation  of the exclusion probability from one. Thus, the setup we study allows for an investigation of the exclusion probability or two-particle exclusion statistics.

{\em Quantum mechanical anyons:} 
%we now present a fully quantum mechanical analysis of the anyon collider. To this end, 
we introduce the operator for tunnelling of a charge $e^\star$ (measured in units of the electron charge) anyon from the upper to the lower edge 
%
%**********************  qp tunneling operator  ****************
\begin{equation}
A(t) \ = \ \zeta e^{i \phi_u(0,t) - i \phi_d(0,t)} \ \ , \ \ \ I_T \ = \ ie^\star \left( A^\dagger - A\right)  .
\label{Eq3} 
\end{equation}
%**************************************************************
%
Here, $\zeta$ is the tunnelling amplitude, and $I_T$ is the operator for the tunnelling current. The boson fields
$\phi_{\alpha}$ with $\alpha = u, d$ obey the equal time commutation relations $[\phi_\alpha(x), \phi_\beta(y)] = i e^\star \pi \delta_{\alpha, \beta} {\rm sign}(x-y)$, and  describe the charge density via $\rho_{\alpha} = \partial_x \phi_{\alpha}/2 \pi$ \cite{Giamarchi_book}. Due to chirality, the $\phi_\alpha$ obey similar relations for equal points in space but different times, $[\phi_\alpha(x,t_1), \phi_\beta(x,t_2)] = i \delta \pi \delta_{\alpha, \beta} {\rm sign}(t_1- t_2)$. However, the parameter $\delta$ governing the time evolution 
may differ from the fractional charge $e^\star$ in the equal time commutator, as a possible consequence of an edge structure with counter-propagating modes coupled locally to  the charge mode that  we focus on \cite{RoHa02}.

In equilibrium, the correlation function of the tunneling operator is given by
%
%*********************  tunneling operator correlation function *****************
\begin{equation}
\langle A(t) A^\dagger(0)\rangle_{\rm eq} \ = \ |\zeta|^2 \, e^{i \pi \delta\,  {\rm sign}(t)} \, {\tau_c^{2 \delta} \over 
|t|^{2 \delta}}  \ .
\label{quantumcorrelation.eq}
\end{equation}
%**********************************************************************************
%
Here, $\tau_c$ denotes a short time cutoff, and $\delta$ may differ from $e^*$ as a consequence of an edge structure with counter-propagating modes coupled locally to  fractional edges \cite{RoHa02}. In order to describe the collision of two dilute anyons beams, 
we need to specify non-equilibrium correlation functions. We assume that the anyons are injected into upper and lower edge via a weak tunnel coupling, such that tunneling events  are rare and uncorrelated in time. Then, we can decompose the boson field according to $\phi_{\alpha} = \phi_{\alpha}^{(0)} + 2 \pi \lambda N_\alpha$ (with $\lambda=1/m$ for a Laughlin state, and $\lambda\neq 1/m$ due to non-universal screening in the presence of edge reconstruction \cite{KaFiPo94,Levkivskyi14}) into an equilibrium part $\phi_\alpha^{(0)}$ describing quantum fluctuations, and a non-equilbrium component $N_\alpha$ describing classical, Poisson distributed fluctuations with expectation value $\langle \dot{N}_\alpha\rangle = \langle I_{\alpha, 0}\rangle/e^\star$. Here, $I_{\alpha, 0}$ denotes the current on edge $\alpha$ before tunneling at the QPC3 takes place. The non-equilibrium contribution
to the correlation function of the tunneling operator is given by the generating function of a Poisson process \cite{Levkivskyi14}.
Tunneling of quasi particles can be considered uncorrelated Poissonian events since the long-time tails in the equilibrium correlation function Eq.~(\ref{quantumcorrelation.eq}) are cut off due to  oscillatory factors 
proportional to the  particle currents $\langle I_{u,0}\rangle/e^*$ and  $ \langle I_{d,0}\rangle/e^*$. Then,   for times $t>0$
%
%*******************   
\begin{eqnarray}
\langle A(t) A^\dagger(0)\rangle_0 & = & \langle A(t) A^\dagger(0)\rangle_{\rm eq}  \hspace*{-.5cm}   \label{classicalcorrelation.eq}
  \\
& & \hspace*{-2.7cm} \times   \exp\!\left[- {\langle I_{u,0}\rangle \over e^\star}\! \left(1 - e^{- 2 \pi i \lambda}\right)\! t \right]  \exp\!\left[
- {\langle I_{d,0}\rangle \over e^\star}\! \left(1 - e^{2 \pi i \lambda}\right)\! t\right]   . \nonumber
\end{eqnarray}
%*********************************
%
For times $t<0$, we have $\langle A(t) A^\dagger(0)\rangle_0 = \langle A(-t) A^\dagger(0)\rangle_0^*$. 
This equation is valid for in the asymptotic regime where the magnitude of the exponents is much larger one, and thus breaks down 
for integer values of $\lambda$.

We are now in a position to compute the expectation values of the tunnelling current and its fluctuations,
%
%******************* general current and tunnelling noise expectation values  *********
\begin{subequations}
\begin{eqnarray}
\hspace*{-0.5cm} \langle I_T \rangle & = & e^\star \int_{-\infty}^\infty \! \! \!   dt \langle [A^\dagger(0), A(t)]\rangle_0 \ ,  \\[.5cm]
\hspace*{-0.5cm}\langle (\delta I_T)^2\rangle_{\omega=0} &= & (e^\star)^2 \int_{-\infty}^\infty  \! \! \! dt \, \langle\{ A^\dagger(0) , A(t)\}\rangle_0  \  . 
\end{eqnarray}
\end{subequations}
%**********************************************************************
%
Here, $[.,.]$ denotes the commutator, and $\{.,.\}$ the anti-commutator. 
Using the correlation functions Eqs.~(\ref{quantumcorrelation.eq}), (\ref{classicalcorrelation.eq}), we obtain for the tunnelling current
%
%**************** tunnel current scaling expression *************
\begin{eqnarray}
\langle I_T\rangle & = & {e^\star \over i} |\zeta|^2  \int_{-\infty}^\infty\! \! \! dt
{\sin\left[{I_- \over e^\star} t \sin 2 \pi \lambda\right] (\tau_c)^{2 \delta}  \over 
\exp\left[{I_+ \over e^\star} t(1 - \cos 2 \pi \lambda)\right] (\tau_c - it)^{2 \delta}} 
\label{tunnelcurrent.eq} \nonumber \\[.5cm]
& = &\! C \sin(\pi \delta) {\rm Im}\left(I_++ {iI_-\over \tan\pi\lambda}\right)^{2\delta-1} \! \!   \left[1 + O(\tau_c)\right] ,
\end{eqnarray}
%********************************************
%
where $O(\tau_c) \to 0 $ for $\tau_c \to 0$, 
$C=e^\star 4 |\zeta|^2 \tau_c^{2 \delta} [\pi(1 - \cos 2 \pi \lambda)/e^\star]^{-1 + 2 \delta} \Gamma(1 - 2 \delta)$, 
$I_+ = |\langle I_{u,0}\rangle|   + |\langle I_{d,0}\rangle| $, and $ I_- = \langle I_{u,0}\rangle   - \langle I_{d,0}\rangle$. 
%******************************************************
%
Similarly, fluctuations in the tunnelling current are given by
%
%***********************  tunnelling current fluctuations *********************
\begin{eqnarray}
{\langle \delta I_T^2\rangle_{\omega=0}\over  (e^\star)^2}
\! & = &\! |\zeta|^2 \!  \int_{-\infty}^\infty\! \! \! \! \! dt {2 \cos\left[{I_- \over e^*} t \sin 2 \pi \lambda\right] 
\tau_c^{2 \delta}  \over \exp\left[{I_+ \over e^\star} t(1 - \cos 2 \pi \lambda)\right]\! \! (\tau_c - it)^{2 \delta}}    \nonumber \\
 &\hspace*{-.9cm} = & \hspace*{-0.5cm} { C \over e^\star} \cos(\pi \delta) {\rm Re}\left(I_++ {iI_-\over\tan\pi\lambda}\right)^{2\delta-1}\! \! \!  \left[1 + O(\tau_c)\right]
. \hspace*{-0.1cm} \nonumber \\ & &  \label{tunnelnoise.eq}
\end{eqnarray}
%*********************************************************************
%

%
%*********************************  full noise minussign   **************
\begin{figure}[t]
\includegraphics[width=\linewidth]{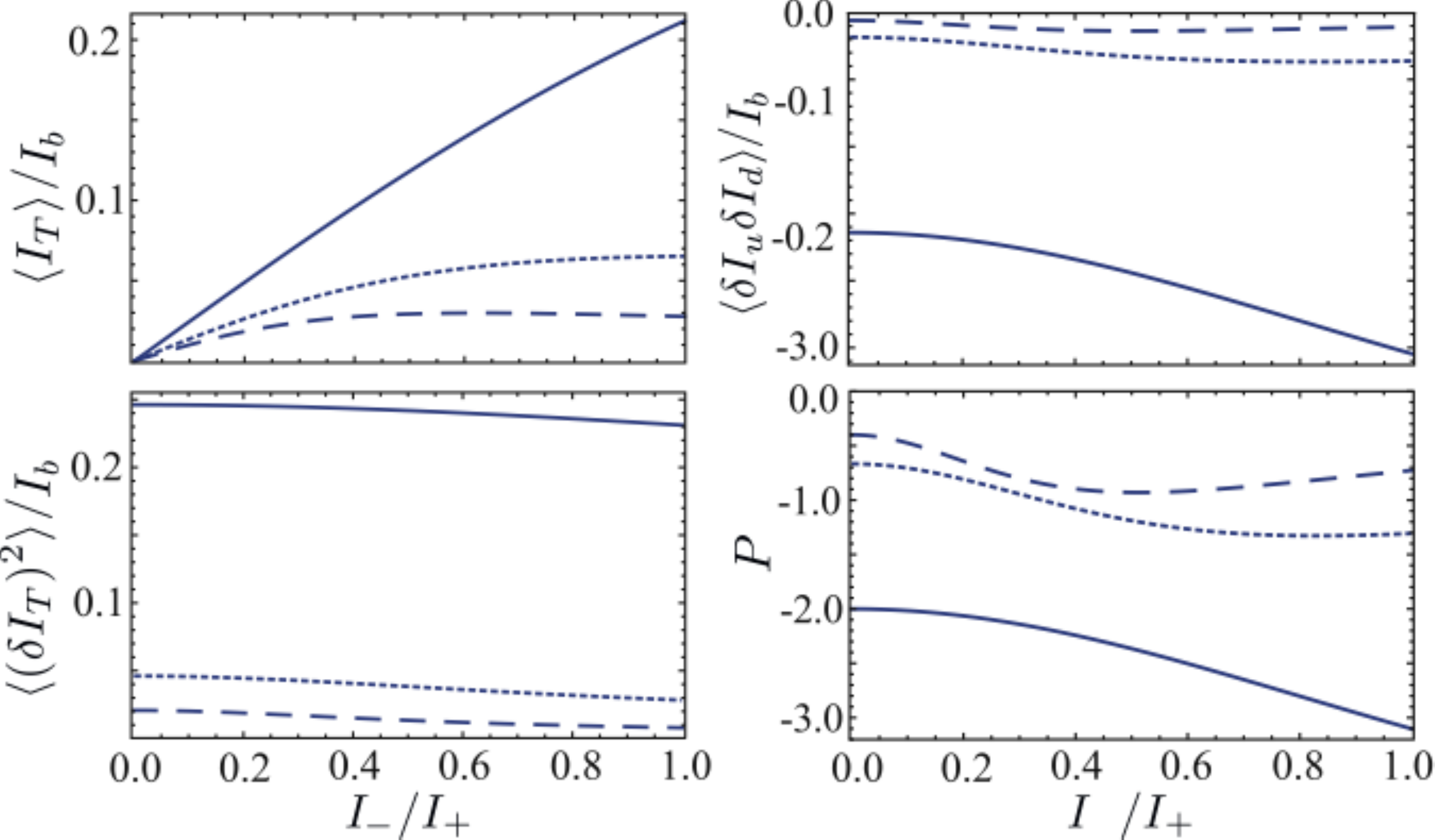}
\caption{Expectation value of the tunnel current (upper left panel) and of its fluctuation strength (lower left panel)
as a function 
of the net bias applied to the middle QPC according to Eqs.~(\ref{tunnelcurrent.eq}) and (\ref{tunnelnoise.eq}). Plot for $e^\star \equiv  \delta \equiv \lambda = 1/m$, 
the full line shows $m = 3$, dotted line $m = 5$, dashed line $m = 7$,  with $I_b=|\zeta|^2 \tau_c^{2 \delta}(I_+)^{-1+2 \delta}$.
Cross-correlations between current fluctuations in upper and lower edge according to Eq.~(\ref{FDT.eq}) (upper right panel) and Eq.~(\ref{noisenormalized.eq}) (lower right panel) as a function of 
the effective bias $I_-/I_+$ between upper and lower edge. Plot for $e^\star \equiv  \delta \equiv \lambda = 1/m$, 
the full line shows $m = 3$, dotted line $m = 5$, dashed line $m = 7$,   with $I_b=|\zeta|^2 \tau_c^{2 \delta}(I_+)^{-1+2 \delta}$. The current cross-correlations have a maximum for zero net bias $I_-/I_+=0$. 
 }
\label{fig:3}
\end{figure}
%***************************************************************
%
These perturbative expressions are of order $|\zeta|^2$ and are valid in the regime $\langle I_T\rangle \ll  \langle I_{u,0}\rangle, \langle I_{d,0}\rangle$. 

In order to compute the current cross-correlation function $\langle \delta I_d \delta I_u \rangle_{\omega=0}$, we parameterize the currents $I_u$ and $I_d$ after the QPC as $I_u = I_{u,0} - I_T$, $I_d = I_{d,0} + I_T$, such that the correlator of current fluctuations is given by $\langle \delta I_d \delta I_u\rangle = - \langle \delta I_T^2\rangle + \langle \delta I_{u,0} \delta I_T\rangle - \langle \delta I_{d,0} \delta I_T\rangle$. 
The correlations between fluctuations in the incoming current and fluctuations in the tunneling current can be expressed in terms of the 
differential conductance of the QPC \cite{suppmat}, such that we obtain 
%
%********************************** FDT  **********************************
\begin{eqnarray}
\langle \delta I_d \delta I_u \rangle_{\omega=0} & = & - \langle \delta I_T^2 \rangle_{\omega=0} 
\label{FDT.eq}
\\
& & + e^\star \left( \langle I_{u,0}\rangle {\partial \over 
\partial \langle I_{u,0}\rangle} - \langle I_{d,0}\rangle {\partial \over \partial \langle I_{d,0}\rangle } \right) \langle I_T\rangle  . 
\nonumber
\end{eqnarray}
%***************************************************************************
%
Here, the first term on the r.h.s.~describes noise generated at the QPC3, whereas the second term describes cross-correlations due to fluctuations in $I_{\alpha,0}$, which are partially transmitted through the QPC3. Eq.~(\ref{FDT.eq}) is a generalization of the fluctuation dissipation theorem to a fully non-equilibrium situation in an interacting system. The second term on the r.h.s.~of Eq.~(\ref{FDT.eq}) is a generalization of Johnson-Nyquist noise to a non-equilibrium situation.

In order to quantify the strength of current cross-correlations by 
a generalized Fano factor, we define
%
%********************************  normalised noise power  **********************
\begin{eqnarray}
P(I_-/I_+) & = & {\langle \delta I_d \delta I_u \rangle_{\omega=0} \over \left. e^\star I_+ {\partial \over \partial I_-} \langle  I_T \rangle \right|_{I_-=0}} \  . \label{noisenormalized.eq}
\end{eqnarray}
%******************************************************************
%
To normalize cross-correlations, we divide them   by the second term on the r.h.s.~of Eq.~(\ref{FDT.eq}), which describe current cross-correlations due to a partial tunneling of fluctuations in the incoming currents. 
Using the results Eqs.~(\ref{tunnelcurrent.eq}), (\ref{tunnelnoise.eq}), (\ref{FDT.eq}), we obtain for the normalized noise power at
zero effective bias ($T_u = T_d$)
%
%*************************  noise power at zero bias  ********************
\begin{equation}
P(0)  =  1 - {\tan \pi \lambda \over \tan \pi \delta} \, {1 \over 1 - 2 \delta} 
\xrightarrow[\lambda={1 \over m}, \delta = {1 \over m}]{} \  {-2 \over m-2} \ \ ,
\label{norm_zerobias.eq}
\end{equation}
%*********************************************************************
%
valid for $\lambda$ not equal to an integer (discussion below Eq.~(\ref{classicalcorrelation.eq})) or half-integer \cite{LeSu09}, and without restriction on $\delta$. 
In order to gain intuition for the 
meaning of the normalised noise power, we compare the quantum mechanical expression with that for the lattice model of particles with an exclusion probability $p$. For the lattice model, the case of vanishing bias $I_-=0$ considered above translates into $T_u = T_d$, and the normalized current cross-correlation is given by [see Eq.\ (\ref{classical_correlation.eq})]  
%
%***************************  zero bias noise power classical model  *****************
\begin{equation}
P_{\rm cl}(T_u = T_d) \ = \ - (1 - p) T_{u} \  . 
\end{equation}
%*****************************************************************
%
This result indeed allows to establish a proportionality between the r.h.s.~of Eq.~(\ref{norm_zerobias.eq}) and the factor $(1-p)$, suggesting that the current cross-correlations indeed give information about quantum statistics. The absence of QPC tunneling probability $T_u$ or $T_d$ in the quantum mechanical expression Eq.~ (\ref{norm_zerobias.eq}) is due to the remarkable universality of the correlation function Eq.~(\ref{classicalcorrelation.eq}), which depends only on 
the injected currents $\langle I_{u/d,0}\rangle$, and not on the voltages $V_{u/d}$ and QPC tunneling probabilities $T_{u/d}$ separately.

We note  that Eq.\ (\ref{norm_zerobias.eq}) does not depend explicitly on the  quasiparticle charge $e^\star$. As well, it is negative, while $P(0)=0$ for free fermions \cite{footn}, even at finite temperature, see Eq.\ (\ref{fermion_correlation.eq}).
Moreover, repulsive Coulomb interactions or edge reconstruction, could only increase $\delta$ and $\lambda$, making $P(0)$ positive for fermions. Thus, if an experiment shows a negative value, then it is a robust evidence of the anyon statistics. 
Among all Laughlin states with odd-integer $m$, expression (\ref{norm_zerobias.eq}) reaches a maximum in magnitude $P(0)=-2$ for $m=3$, and then monotonically decreases in magnitude towards zero, which is reached in the hypothetical limit $m\to \infty$. The fact that $P(0)<-1$ for $m=3$ suggests $p<0$ or bunching for anyons, so that tunneling to an edge is more favorable when another anyon is present at the tunneling point.

Why do the normalized cross-correlations Eq.~(\ref{norm_zerobias.eq}) vanish in the limit of large $m$?  One can check that in 
a thermal state cross-correlations vanish, $P_{\rm thermal}(0) \equiv 0$, because the two contributions on the r.h.s.~of Eq.~(\ref{FDT.eq}) exactly cancel each other. 
When computing tunneling current and noise according to Eqs.~(\ref{tunnelcurrent.eq}), (\ref{tunnelnoise.eq}) at zero effective bias $I_-=0$, the non-equilibrium correlation function $\exp(- 2 \pi \Theta |t|/m)/ |t|^{2/m}$ with $\Theta = I_+(1-\cos 2\pi\lambda)/2\pi$ in general differs from the thermal correlation function at temperature $\Theta$, that reads $ 1/\sinh(\pi \Theta t)^{2/m}$. 
However, in the integrand of Eqs.~(\ref{tunnelcurrent.eq}), (\ref{tunnelnoise.eq}), the non-equilbrium correlation function leads to the same result in the limit
of large $m$ as the thermal correlation function, explaining the disappearance of cross-correlations in this limit. We thus interpret 
the smallness of cross-correlations in the limit of large $m$ as an indication that the non-equilibrium steady state is similar to a thermal state.

In summary, we have analyzed a setup in which two dilute beams of anyons collide with each other at a QPC. Correlations of current fluctuations after the QPC are a measure of how strongly particles exclude each other spatially. We find that the absence of current correlations for fermions is due to perfect spatial exclusion, while the negative correlations for anyons are a clear signature of a reduced spatial exclusion as compared to fermions. On a qualitative level, the results of our quantum mechanical non-equilibrium calculation can be described within the framework of a classical lattice model
with partial exclusion. 
The quantum results are independent of the short-range interaction
between anyons and, in contrast to the classical model, depend only on the
total injected currents $\langle I_{u/d,0} \rangle$ and not on the voltage V or the
transmission probabilities $T_{u/d}$ individually. The results are
governed by parameters $\lambda$ and $\delta$,  which are equal to the bulk statistical parameter in the simplest case, but which may be renormalized by interaction with backward moving modes in the event of edge reconstruction.

We would like to thank I.~Gurman, M.~Heiblum,  R.~Sabok, and  E.~Sukhorukov for stimulating discussions. We would like to acknowledge support by DFG grants RO 2247/7-1 and RO 2247/8-1, from the Microsoft
Corporation, and from Swiss NSF.

\appendix
\widetext
\vspace{1cm}
\begin{center}
\large{\bf Supplemental material}
\end{center}

\section{Derivation of the cross-correlations in the classical model}
We study a system of two lines of particles arriving at a QPC, where the probability to end up after the QPC on the same side is $2(1-p)T(1-T)$. The number of particles passing a given point in each line within a fixed time window $\Delta t$, $N_{u/d, 0}$, fluctuates with the probability of occupation of each site $T_{u}$ or $T_d$ correspondingly. We then calculate the average $\langle \delta N_u\delta N_d\rangle$ for the fluctuations of the numbers of transmitted particles in two steps using conditional parameters $N_{u,0}$, $N_{d,0}$, $N_c$.
\begin{subequations}
\begin{eqnarray}
N_u = N_{u,0} - N_T - \Delta N_c, \\
N_d = N_{d,0} + N_T + \Delta N_c,
\end{eqnarray}
\end{subequations}
where $N_T$ is the number of particles tunneled due to non-coincident events, and $\Delta N_c$ the number tunneled due to coincident events. The variables have the following statistics, $\langle N_{u/d,0}\rangle = T_{u/d}V$ and
\begin{equation}
\langle N_{u/d,0}^2\rangle = \langle \delta N_{u/d,0}^2\rangle +\langle N_{u/d,0}\rangle^2 = T_{u/d}(1-T_{u/d})V +T_{u/d}^2V^2 \  ,
\label{2cond}
\end{equation}
and the average number of coincidents is given by 
\begin{equation}
\langle N_c\rangle = T_uT_dV  \  .
\end{equation}
For the tunneled numbers we have $\langle N_T\rangle = T(N_{u,0}-N_{d,0})$, $\langle\Delta N_c\rangle = 0$, and
\begin{eqnarray}
\langle N_T^2\rangle = \langle \delta N_T^2\rangle + \langle N_T\rangle^2 &=& T(1-T)(N_{u,0}+N_{d,0}-2N_c)+T^2(N_{u,0}-N_{d,0}) \  , \\
\langle \Delta N_c^2\rangle &=& 2(1-p)N_c T(1-T) \  .
\end{eqnarray}
Next, we calculate the conditional average using the two last equations, this leads to
\begin{equation}
\langle N_u N_d\rangle_{\rm cond} = -T(1-T)(N_{u,0}+N_{d,0} - 2pN_c) +T(1-T)(N_{u,0}-N_{d,0})^2 + N_{u,0}N_{d,0} \  .
\end{equation}
Then, we average over the conditional parameters using Eq.\ (\ref{2cond}) and obtain
\begin{equation}
\langle N_u N_d\rangle = -T(1-T)V(T_u+T_d - 2pT_uT_d) + T(1-T)V[T_u(1-T_u) + VT_u^2 + T_d(1-T_d) + VT_d^2-2VT_uT_d] +V^2T_uT_d \  .
\end{equation}
The average numbers are given by
\begin{subequations}
\begin{eqnarray}
\langle N_u\rangle = T_uV + TV(T_d-T_u),\\
\langle N_d\rangle = T_dV - TV(T_d-T_u),
\end{eqnarray}
\end{subequations}
so that, finally, the correlator is given by
\begin{equation}
\langle \delta N_u\delta N_d\rangle = \langle N_u N_d\rangle - \langle N_u\rangle \langle N_d\rangle = -T(1-T)V[(T_u-T_d)^2 - 2(1-p) T_uT_d].
\end{equation}
Note that the first term vanishes for $T_u = T_d$, while the second term is zero for fermions with $p=1$.
Also we note that for  particles with uncorrelated tunnelling,  $p=0$, there is no cross-term $\propto T_u T_d$ in 
 $\langle \delta N_u\delta N_d\rangle = -T(1-T)V[T_u^2+T_d^2]$.

\section{Quantum derivation of the current cross-correlations}
%
%***************  expression cross-correlations  ***********
In this section we consider the cross-terms in Eq.\ (\ref{FDT.eq}) and evaluate them by using the non-equilibrium bosonization technique. Since both terms are similar, we focus on one of them, namely
\begin{equation}
\langle \delta I_T \delta  I_{u,0} \rangle   =  \int_{-\infty}^\infty dt \langle I_T(0) \delta I_{u,0}(t) \rangle =    - i e^\star  \int_{-\infty}^\infty  dt \int_{-\infty}^0 dt^\prime \langle [ A(t^\prime)+ A^\dagger(t^\prime), A(0) - A^\dagger(0)] {v \over 2 \pi} \partial_x \delta \phi_u(t) \rangle,
\end{equation}
where we used the second equation in (\ref{Eq3}) for the tunneling current, and the bosonic expression for the current at the edge, see Ref. \cite{Giamarchi_book}. Next, we use the first equation in (\ref{Eq3}) for the tunneling amplitude and find that
\begin{eqnarray}
\langle \delta I_T \delta  I_{u,0} \rangle & = & -i e^\star {v \over 2 \pi} \int_{-\infty}^\infty  dt \int_{-\infty}^0 dt^\prime\nonumber
\\ & &\left[ i G_d^<(-t^\prime) \langle e^{i \phi_u(t^\prime) - i \phi_u(0)} \partial_x \delta \phi_u(t) \rangle  e^{i {\pi \over 2m} {\rm sign}(t^\prime)}
+ 
i G_d^>(t^\prime) \langle e^{i \phi_u(0) - i \phi_u(t^\prime)} \partial_x \delta \phi_u(t)\rangle  e^{i {\pi \over 2m} {\rm sign}(t^\prime)}\right.
\label{updowncorrelator.eq}\\
& & \left. +i G_d^<(t^\prime)\langle e^{i \phi_u(0) - i \delta \phi_u(t^\prime)} \partial_x \delta \phi_u(t)\rangle e^{-i {\pi \over 2m} {\rm sign}(t^\prime)}
+ i G_d^>(- t^\prime) \langle e^{i \phi_u(t^\prime) - i \phi_u(0)} \partial_x \delta \phi_u(t) \rangle e^{-i {\pi \over 2m} {\rm sign}(t^\prime)}\nonumber 
\right]
\end{eqnarray}
%**********************************
%
where the Green functions are defined as
%
%******************  lesser and greater Green functions *********
\begin{subequations}
\begin{eqnarray}
i G_d^>(t) & = & \langle e^{i \phi_d(t)} e^{-i \phi_d(0)}\rangle \  , \\[.5cm]
-i G_d^<(t) & = & \langle e^{-i \phi_d(0)} e^{i \phi_d(t)}\rangle \  .
\end{eqnarray}
\end{subequations}
%********************************
%
Here,  we have decomposed the boson field at the upper channel into equilibrium and non-equilibrium parts $\delta \phi_u = \phi_u^{(0)} + {2 \pi \over m} \delta N_u$, 
and introduce the notations $\delta j_u \equiv  e^\star \delta \dot{N}_u$, and  $\delta N_u \equiv N_u   - \langle N_u\rangle$. 
Here, $\delta j_u$ denotes non-equilibrium fluctuations of the current in the upper wire, whereas $\delta I_u$ denotes fluctuations in the total current.  We now focus on one of the four terms in Eq. (\ref{updowncorrelator.eq}), say the first one.
Since the equilibrium part does not contribute to (\ref{updowncorrelator.eq}) we find
%
%*********************************   decomposition of expectation values ******
\begin{equation}
\int_{-\infty}^\infty dt \langle e^{i \phi_u(t^\prime) - i \phi_u(0)} {v\over 2 \pi} \partial_x \delta \phi_u(t)\rangle 
 =  \int_{-\infty}^\infty dt \langle e^{i {2 \pi \over m}\left[N_u(t^\prime) - N_u(0)\right]} \delta j_u(t)\rangle
 {1 \over |t^\prime|^{1/m}}  \ .
 \end{equation}
%*************************
%
Due to the fact that the main contribution to the integral in Eq.\ (\ref{updowncorrelator.eq}) comes from long times, the current fluctuations can be considered Markovian, i.e., $\delta j_u(t)$ has short-range correlations in time. It is clear then that the $t$-integral in above equation can only contribute in the interval $[t^\prime, 0]$. 
 \begin{eqnarray} \label{supleq1}
\int_{-\infty}^\infty dt \langle e^{i \phi_u(t^\prime) - i \phi_u(0)} {v\over 2 \pi} \partial_x \delta \phi_u(t)\rangle & = & \int_{t^\prime}^0 dt  \langle e^{i {2 \pi \over m}\left[N_u(t^\prime) - N_u(0)\right]} \delta j_u(t)\rangle
 {1 \over |t^\prime|^{1/m}}\\[.5cm]
 & = & - e^{i {2 \pi \over m}[\langle N_u(t^\prime) \rangle - \langle N_u(0)\rangle]} \langle e^{i {2 \pi \over m} [\delta N_u(t^\prime) - \delta N_u(0)]} \left[\delta N_u(t^\prime) - \delta N_u(0)\right]e^\star \rangle {1 \over |t^\prime|^{1/m}} \nonumber \  .
\end{eqnarray}
We now express the multiplicative factor as a derivative of the exponential with respect to a "counting variable" $\lambda$ 
%
%*******************  
\begin{equation}
\langle e^{i {2 \pi \over m} [\delta N_u(t^\prime) - \delta N_u(0)]} \left[\delta N_u(t^\prime) - \delta N_u(0)\right]\rangle  =  {\partial \over i \partial \lambda}\left.  \langle  e^{i \lambda [\delta N_u(t^\prime) - \delta N_u(0)]}\rangle  
\right|_{\lambda = {2 \pi \over m}} =   {\partial \over i \partial \lambda}\left. e^{  - t^\prime  {\langle I_{u,0} \rangle \over e^\star} \left(e^{-i \lambda} - 1 + i \lambda\right)}  \right|_{\lambda = {2 \pi \over m}}
\  .
\end{equation}
Finally, we find that
\begin{equation}
\langle e^{i {2 \pi \over m} [\delta N_u(t^\prime) - \delta N_u(0)]} \left[\delta N_u(t^\prime) - \delta N_u(0)\right]\rangle= - t^\prime {\langle I_{u,0} \rangle \over e^*}  e^{  - t^\prime  {\langle I_{u,0} \rangle \over e^\star} \left(e^{-2\pi i/m } - 1 + 2\pi i /m\right)}
(-e^{- 2\pi i /m} + 1)
\label{supleq2} \  .
\end{equation}
%*****************************************
%
Here, we took the complex conjugate of the correlation function Eq.~(\ref{classicalcorrelation.eq}) and changed the overall sign of the exponent because we used the correlation function for negative times $t^\prime < 0$.  We now use the fact that $\langle N_u(t) \rangle = \langle I_{u,0} \rangle t/e^\star$, and thus obtain  
%
%******************  non equilibrium cross-correlation function ****************
\begin{eqnarray}
e^{i {2 \pi \over m}[\langle N_u(t^\prime) \rangle - \langle N_u(0)\rangle]} 
&=& e^{i {2 \pi \over m} {\langle I_{u,0} \rangle \over e^\star} t^\prime}  \  .
\end{eqnarray}
%******************************************************************
%
Combining the results (\ref{supleq1}) and (\ref{supleq2}), we find that 
%
%***************** non equilibrium cross-correlation function final expression ***********
\begin{eqnarray}
\int_{-\infty}^\infty dt \langle e^{i \phi_u(t^\prime) - i \phi_u(0)} {v\over 2 \pi} \partial_x \phi_u(t)\rangle
& = & t^\prime {\langle I_{u,0} \rangle \over e^*}  e^{  - t^\prime  {\langle I_{u,0} \rangle \over e^\star} \left(e^{-i {2 \pi \over m}} - 1\right)}
(-e^{- i {2 \pi \over m}} + 1) {e^\star  \over |t^\prime|^{1/m}}\\[.5cm]
& = & e^\star \langle I_{u,0} \rangle {\partial \over \partial \langle I_{u,0} \rangle }
e^{  - t^\prime  {\langle I_{u,0} \rangle \over e^\star} \left(e^{-i {2 \pi \over m}} - 1\right)}  \ .
\end{eqnarray}
%*************************************
%
We note that this relation holds for all four terms in Eq.~(\ref{updowncorrelator.eq}). In addition, the 
expectation value of the tunnelling current is given by
%
%************ abstract Expression tunnelling current  *************
\begin{eqnarray}
\langle I_T \rangle & = & - i e^\star  \int_{-\infty}^\infty  dt \int_{-\infty}^0 dt^\prime \langle [ A(t^\prime)+ A^\dagger(t^\prime), A(0) - A^\dagger(0)]  \rangle \ \ , 
\end{eqnarray}
%*************************************************************
%
and we thus obtain the relation 
%
%****************** FDT  *************
\begin{eqnarray}
\langle \delta I_T \delta I_{u,0}\rangle  & = & e^\star \langle I_{u,0} \rangle {\partial \over \partial \langle I_{u,0} \rangle } 
\langle I_T \rangle   \  .
\end{eqnarray}
%*********************************
%
Similarly, we find 
%
%****************** FDT  *************
\begin{eqnarray}
\langle \delta I_T \delta I_{d,0} \rangle & = & e^\star \langle I_{d,0} \rangle {\partial \over \partial \langle I_{d,0} \rangle } 
\langle I_T \rangle  \  .
\end{eqnarray}
%*********************************
%


\begin{thebibliography}{99}
\bibitem{HBT54} R.~Hanbury Brown and R.Q.~Twiss, Philos. Mag. {\bf 45}, 663 (1954).
\bibitem{HBT56} R.~Hanbury Brown and R.Q.~Twiss, Nature  {\bf 177}, 27 (1956).
\bibitem{Neder+07}  I.~Neder, N.~Ofek, Y.~Chung, M.~Heiblum, D.~Mahalu, and V.~Umansky, Nature  {\bf 448}, 333 (2007).
\bibitem{LeMy77}J.M.~ Leinaas and J.~Myrheim,   Il Nuovo Cimento B {\bf 37}, 1 (1977).
\bibitem{Laughlin83} R.B.~Laughlin, Phys. Rev. Lett. {\bf 50}, 1395  (1983).
\bibitem{Halperin84} B.I.~Halperin, Phys. Rev. Lett. {\bf 52} , 1583 (1984).
\bibitem{ArScWi84} D.~Arovas, J.R.~Schrieffer, and F.~Wilczek,  Phys. Rev. Lett. {\bf 53}, 722  (1984).

\bibitem{Campagnano+12} G.~Campagnano, O.~Zilberberg, I.V.~Gornyi,  D.E.~Feldman,  A.C.~Potter, and Y.~Gefen, Phys. Rev. Lett. {\bf 109}, 106802 (2012).
\bibitem{Halperin82} B.I.~Halperin, Phys. Rev. B {\bf  25}, 2185  (1982). 
\bibitem{Wen90} X.-G.~Wen, Phys. Rev. B {\bf 41},12838 (1990).
\bibitem{Landauer70} R.~Landauer, Philos. Mag. {\bf 21}, 863 (1970).
\bibitem{Buttiker88} M.~Buttiker, Phys. Rev. B {\bf 38},    9375   (1988). 
\bibitem{footn} Note that limit $m=1$ should not be taken in Eq.\ (\ref{norm_zerobias.eq}) since Eq.\ (\ref{classicalcorrelation.eq}) is not valid at this point.
\bibitem{KaFi03} C.L.~Kane and M.P.A.~Fisher, Phys. Rev. B {\bf 67}, 045307 (2003).
\bibitem{LeSu09} I.P. Levkivskyi and E.V. Sukhorukov, Phys. Rev. Lett. \textbf{103}, 036801 (2009);
Phys. Rev. B {\bf 85}, 075309 (2012).
\bibitem{GuGeMi} D. B. Gutman, Y. Gefen and A.D. Mirlin, Phys. Rev. B \textbf{81} 085436 (2010); J. Phys. A:Math. Theor. \textbf{44}, 165003 (2011)
\bibitem{Levkivskyi14}  I.P.~Levkivskyi, preprint  arXiv:1402.1989 (2014). 
\bibitem{BlBu00} Y.M.~Blanter and M.~Buttiker, Phys. Rep. {\bf 336}, 1 (2000). 
\bibitem{Giamarchi_book} see, e.g., T.~Giamarchi, {\em Quantum Physics in One Dimension}, Oxford Univ. Press (2004). 
\bibitem{RoHa02} B.~Rosenow and B.I.~Halperin, Phys. Rev. Lett. {\bf 88}, 096404 (2002).
\bibitem{KaFiPo94} C.L.~Kane, M.P.A.~Fisher, and J.~Polchinski, Phys. Rev. Lett. {\bf 72}, 4129 (1994). 
\bibitem{suppmat} see supplemental material.
\end{thebibliography}
\end{document}